\documentclass[12pt,preprint]{aastex}
\usepackage{epsfig}
\usepackage{graphicx}
\usepackage{rotating}
\usepackage{slashbox}
\usepackage{multirow}
\usepackage{color}
\usepackage{lscape}
\usepackage{mathrsfs,amssymb}
\usepackage{amsmath}
\usepackage{morefloats}

\newcommand	\beq	{\begin{equation}}	
\newcommand	\eeq	{\end{equation}}	
\newcommand       \Angstrom     {\,{\rm \AA}}

\newcommand       \nm           {\,{\rm nm}}
\newcommand       \cm           {\,{\rm cm}}

\newcommand       \eV           {\,{\rm eV}}
\newcommand       \g            {\,{\rm g}}
\newcommand       \K            {\,{\rm K}}
\newcommand       \pc           {\,{\rm pc}}

\newcommand       \s            {\,{\rm s}}

\newcommand       \yr       {\,{\rm yr}}

\newcommand       \simlt        {\lesssim}
\newcommand       \simgt        {\gtrsim}
\newcommand       \gtsim        {\gtrsim}

\newcommand       \mum          {\,{\rm \mu m}}

\newcommand       \Tpeak        {T_{\rm peak}}
\newcommand       \TD           {T_{\rm D}}
\newcommand       \Natom        {N_{\rm atom}}
\newcommand       \Cabs         {C_{\rm abs}}

\newcommand       \kabsnano     {\kappa_{\rm abs}^{\rm nano}}

\newcommand       \simali       {\sim\,}

\newcommand{\spitzerirs}{{\em Spitzer}/IRS\ }

\newcommand{\etal}{\textrm{et al.\ }}

\newcommand{\eg}{\textrm{e.g., }}

\newcommand{\Msun}      {\,{ M_\odot}}

\def\lax{{$\mathrel{\hbox{\rlap{\hbox{\lower4pt\hbox{$\sim$}}}\hbox{$<$}}}$}}
\def\gax{{$\mathrel{\hbox{\rlap{\hbox{\lower4pt\hbox{$\sim$}}}\hbox{$>$}}}$}}

\countdef\decade=200
\advance\decade by \year
\countdef\hours=201
\advance\hours by \time
\divide\hours by 60
\countdef\mins=202
\advance\mins by \hours
\multiply\mins by 60
\multiply\hours by 100
\countdef\miltime=203
\advance\miltime by \hours
\advance\miltime by \time
\advance\miltime by -\mins

\begin{document}
\title{The Widespread Presence of Nanometer-size Dust Grains in the Interstellar Medium of Galaxies}
\author{
Yanxia~Xie\altaffilmark{1},  
Luis~C. Ho\altaffilmark{1,2}, 
Aigen~Li\altaffilmark{3}, and 
Jinyi~Shangguan\altaffilmark{1,2} 
}
\altaffiltext{1}{Kavli Institute for Astronomy and Astrophysics, 
                       Peking University, Beijing 100871, China;  
                       {\sf yanxia.xie@pku.edu.cn, lho.pku@gmail.com}
                       }
\altaffiltext{2}{Department of Astronomy, School of Physics, 
                      Peking University, Beijing 100871, China; 
                       {\sf shangguan@pku.edu.cn}
                       }                       
\altaffiltext{3}{Department of Physics and Astronomy, 
                       University of Missouri, 
                       Columbia, MO 65211, USA;
                       {\sf lia@missouri.edu}
                       }

\begin{abstract}
Interstellar dust spans a wide range in size distribution, ranging
from ultrasmall grains of a few $\rm \AA$ngstr\"oms to
micronmeter-size grains. While the presence of nanometer-size dust
grains in the Galactic interstellar medium was speculated six decades
ago and was previously suggested based on early infrared observations,
systematic and direct analysis of their properties over a wide range
of environments has been lacking. Here we report the detection of
nanometer-size dust grains that appear to be universally present in a
wide variety of astronomical environments, from Galactic high-latitude
clouds to nearby star-forming galaxies and galaxies with low levels of
nuclear activity. The prevalence of such a grain population is
revealed conclusively as prominent mid-infrared continuum emission at
$\lambda$\,$\simlt$\,10$\mum$ seen in the {\it{Spitzer}}/IRS data,
characterized by temperatures of $\simali$300--400$\K$ that are
significantly higher than the equilibrium temperatures of common,
submicron-size grains in typical galactic environments.  We propose
that the optimal carriers of this pervasive, featureless hot dust component are
very small carbonaceous (e.g., graphite) grains of nanometer size that are transiently heated by single-photon absorption.  This grain population accounts for $\simali$1.4\% of the total infrared emission at $\simali$5--3000$\mum$ and $\simali$0.4\% of the total interstellar dust mass. 

\end{abstract}


\section{Introduction\label{sec:intro}}
The concept of very small ($\simali$10$\Angstrom$) dust grains was first put forward to account for interstellar reddening (Platt 1956). It was later realized that these grains are small enough that their time-averaged heat capacities are comparable to or smaller than the energy of a single photon that heats the grains (Greenberg 1968). Theoretical calculations predict that the temperature of ultrasmall grains suffers large fluctuations and can be stochastically heated by absorbing a single photon to reach instantaneously several hundred Kelvin, considerably higher than the equilibrium temperature of $\simali$15--30\,K reached with the radiation field for submicron grains (Duley 1973; Greenberg \& Hong 1974; Purcell 1976; Andriesse 1978; Draine \& Anderson 1985).
Exposed to starlight of different levels of intensity (\eg at
different distances from the star), stochastically heated
nanometer-size grains emit in the infrared (IR) essentially with the same
spectral profile characterized by the same temperature (Sellgren 1984;
Draine \& Li 2001).

Sporadic evidence for the existence of very small grains came from
early observations of Galactic and extragalactic sources. The first
detection was reported in the H\,II region M\,17\,S, which showed very
similar source profiles in the IR characterized by a
temperature of $\simali$150$\K$, independent of the distance from the
heating star (Andriesse \& Vries 1976). Similarly, in three visual
reflection nebulae, a uniform near-IR continuum emission at
$\simali$2--5$\mum$ was detected to have a color temperature of
$\simali$800--1000$\K$, again independent of distance from the star
(Sellgren \etal 1983). Two of these three reflection nebulae, as well
as many other environments, also exhibit a broad, featureless
band at $\simali$5400--9500$\Angstrom$, known as the ``extended red
emission'', generally attributed to photoluminescence by nanoparticles
(Witt 2000). A constant continuum emission between $\simali$3--5$\mum$
has been detected in star-forming galaxies based on spectroscopic
observations from the \textit{ Infrared Space Observatory} (ISO); it
appears to depend very weakly on the galaxy parameters and far-IR
colors (Helou \etal 2000; Sturm \etal 2000; Lu \etal 2003). Evidence
for small grains in the Galactic diffuse interstellar medium (ISM) is
also revealed as excess emission from IR cirrus observed at 12 and
25$\mum$ by the {\it Infrared Astronomical Satellite} (IRAS; Boulanger
\& Perault 1988) and at 3.5, 4.9, 12 and 25$\mum$ by the
  Diffuse Infrared Background Experiment (DIRBE) on board the {\it
    Cosmic Background Explorer} (COBE; Arendt \etal 1998).
Furthermore, the steeply rising far-UV extinction at
$\lambda^{-1}>6\mum^{-1}$ (see Li 2004a) and the ``anomalous microwave
emission'', an important Galactic foreground of the cosmic microwave
background radiation in the $\simali$10--100\,GHz region, suggest that
the ISM contains a considerable amount of small grains (see Draine \&
Lazarian 1998, Hensley \& Draine 2017a).

In this work we report the detection of hot ($\simali$300--400$\K$)
continuum emission, which we attribute to nanometer-size dust grains,
over a wide range of Galactic and extragalactic environments. This
component of the ISM, long hypothesized from models and implied in
early observations, is here detected unambiguously in modern
high-quality mid-IR spectra (see \S\ref{sec:obs}).
We propose that nanometer-size carbonaceous grains,  transiently
heated by single-photon absorption, are optimal carriers for this hot
dust component (see \S\ref{sec:size} and \S\ref{sec:composition}). We
show that nanometer-size grains lock up $\simali$1.5\% of the carbon
and $\simali$0.4\% of the total interstellar dust mass (see \S\ref{sec:mass}).

\section{Observational Material and Method of Analysis\label{sec:obs}}

The data analyzed in this work are based on mid-IR spectra taken with
the Infrared Spectrograph (IRS; Houck \etal 2004) onboard the
\textit{Spitzer Space Telescope} (Werner \etal 2004). The IRS has both
a low-resolution and a high-resolution mode. We focus on the
low-resolution spectra, whose broader spectral coverage provides a
better probe of the hot continuum emission. The short-low mode has a
slit size of $3.7^{\prime\prime}\times 57^{\prime\prime}$ and
$3.6^{\prime\prime} \times 57^{\prime\prime}$, covering, respectively,
$\simali$7.4--14.5$\mum$ and $\simali$5.2--7.7$\mum$. The long-low
mode has a slit size of $10.7^{\prime\prime}\times 168^{\prime\prime}$
and  $10.5^{\prime\prime}\times 168^{\prime\prime}$, covering,
respectively, $\simali$19.5--38.0$\mum$ and
$\simali$14.0--21.3$\mum$. The resolution varies from $\simali$64 to
$\simali$128 in each segment. We use the spectra of high
signal-to-noise ratio from the {\it Spitzer Infrared Nearby Galaxy
  Survey} (SINGS\footnote{%
The reduced data are taken from 
the SINGS Legacy Survey web site 
{\tt http://irsa.ipac.caltech.edu/data/SPITZER/SINGS/}.
}; Kennicutt \etal 2003, Smith \etal 2007),
focusing on a sample of 46 objects, comprising 24 star-forming (H~II)
galaxies and 22 galaxies with relatively low-luminosity active
galactic nuclei, including 14 low-ionization nuclear emission-line
regions (LINERs) and eight Seyferts (Ho 2008). In addition to the
galaxy sample, we also choose the IRS low-resolution spectra taken at
four different positions (A--D) within the Galactic HLC
DCld~300.2$-$16.9 (Ingalls \etal 2011), which lies at a distance of $70\pm15\pc$ from the solar neighborhood.

The mid-IR spectra of the ISM in our Galaxy and in external galaxies
exhibit a plethora of emission components from dust and gas. Apart
from the prominent polycyclic aromatic hydrocarbon (PAH) features
located at 6.2, 7.7, 8.6, and 11.3$\mum$, the mid-IR region also
captures dust continuum emission of various temperatures arising from
grains of different sizes, as well as a wealth of ionic and molecular
gas emission lines. The detailed elaboration over the characteristics
of the mid-IR spectra is presented in a companion paper (Xie \etal 2018), which presents a new methodology to decompose the different spectral components. Here we summarize the main steps of spectral decomposition.

To reduce the complexity and potential contamination from many
overlapping components of the mid-IR spectra, we begin by first
removing the ionic and molecular lines that are not blended with PAH
features.  We describe the PAH emission using a theoretical PAH
spectrum calculated adopting a starlight intensity $U$\,=\,1 times of
the interstellar radiation field (ISRF) of 
the solar neighborhood ISM 
(Mathis \etal 1983; hereafter MMP83) 
and grain sizes $3.5\Angstrom < a < 20\Angstrom$. 
The PAH spectrum is relatively insensitive to radiation
intensities ranging up to $U \approx 10^4$.  To account for the dust
continuum emission, we adopt a linear combination of three modified
blackbodies (i.e., graybodies) of different temperatures (``hot'',
``warm'', and ``cold'').\footnote{%
By ``hot'', ``warm,'' and ``cold,'' we merely refer to the different dust components adopted in our methodology to account for the continuum emission at $\simali$5--40$\mum$
underlying the PAH emission features; they differ from the
conventional terminology for dust temperatures in galaxies (Li
2004b).  For example, the conventional terminology defines cold dust as
submicron-size dust of equilibrium temperatures $\simali$15--30$\K$,
which dominates the emission at $\lambda$\,$\simgt$\,100$\mum$, while
the cold component defined here is much warmer ($\simali$40--70$\K$).
  }
All four components are subject to extinction by dust.  In all, the rest-frame spectrum is fit through $\chi^2$ minimization to the following model spectrum:
\begin{equation}
F_\nu =
\left [A^{\rm PAH} J^{\rm PAH}_{\nu} +
    \it A^{h}B_{\nu}(T^{h})/{\lambda}^{\beta} + 
        A^{w}B_{\nu}(T^{w})/{\lambda}^{\beta} + 
        A^{c}B_{\nu}(T^{c})/{\lambda}^{\beta} \right]
     \exp\left(-A_\lambda/1.086\right), 
\label{eq:num1}     
\end{equation}
where $A^{\rm PAH}$ is the scale factor of the PAH template $J^{\rm
  PAH}_{\nu}$; $A^{h}$, $A^{w}$, and $A^{c}$ are, respectively, the
amplitudes of the hot, warm, and cold dust components represented as
modified blackbodies of temperature $T^{h}$, $T^{w}$, and
$T^{c}$; $B_{\nu}(T)$ is the Planck function of temperature $T$ at
frequency $\nu$; $\beta$ is the dust emissivity index, assumed the
same (Li 2009) for all three modified blackbodies and fixed to 2;\footnote{%
Assuming that all dust particles are 
spherical and are composed of electrons and ions, they can be treated as 
a sum of dipoles of classical Lorentz harmonic oscillators, oscillating under the 
force of an electromagnetic field. For submicron-sized (and smaller) grains
that satisfy the Rayleigh limit in the IR ($2\,\pi\,a/\lambda \ll 
1$), the absorption by dust of incident radiation can be approximated as 
$\simali$$\lambda^{-2}$ (i.e., $\beta\approx2$), in accordance with the solution 
for the motion of harmonic oscillators, 
where $\lambda$ is the wavelength of the incident 
electromagnetic wave. In practical situations, 
$\beta$ may vary from $\simali$1.5 to $\simali$2,
which indicates that the vibration of the dust grain deviates from that of 
ideal harmonic oscillation (see Draine 2003 and Li 2009 for more details). 
The determination of the exact value of $\beta$ requires full information on 
the IR SED. In this work, we fix $\beta$ to 2.}
and $A_{\lambda}$ is the extinction, adopted from the theoretical
curve of (Wang \etal 2015). We obtain the best fit for each galaxy using 
the IDL code {\tt MPFIT}, a $\chi^{2}$-minimization routine based on the 
Levenberg-Marquardt algorithm (Markwardt 2009). The error of each parameter 
is estimated based on Monte Carlo simulations (Xie \etal 2018). 
Table~\ref{tab:basic_T} list the derived temperatures and errors of each component 
for the sample.

\section{Results and Discussions\label{sec:results}}
\begin{figure}  
\begin{center}
\resizebox{0.8\hsize}{!}
{\includegraphics{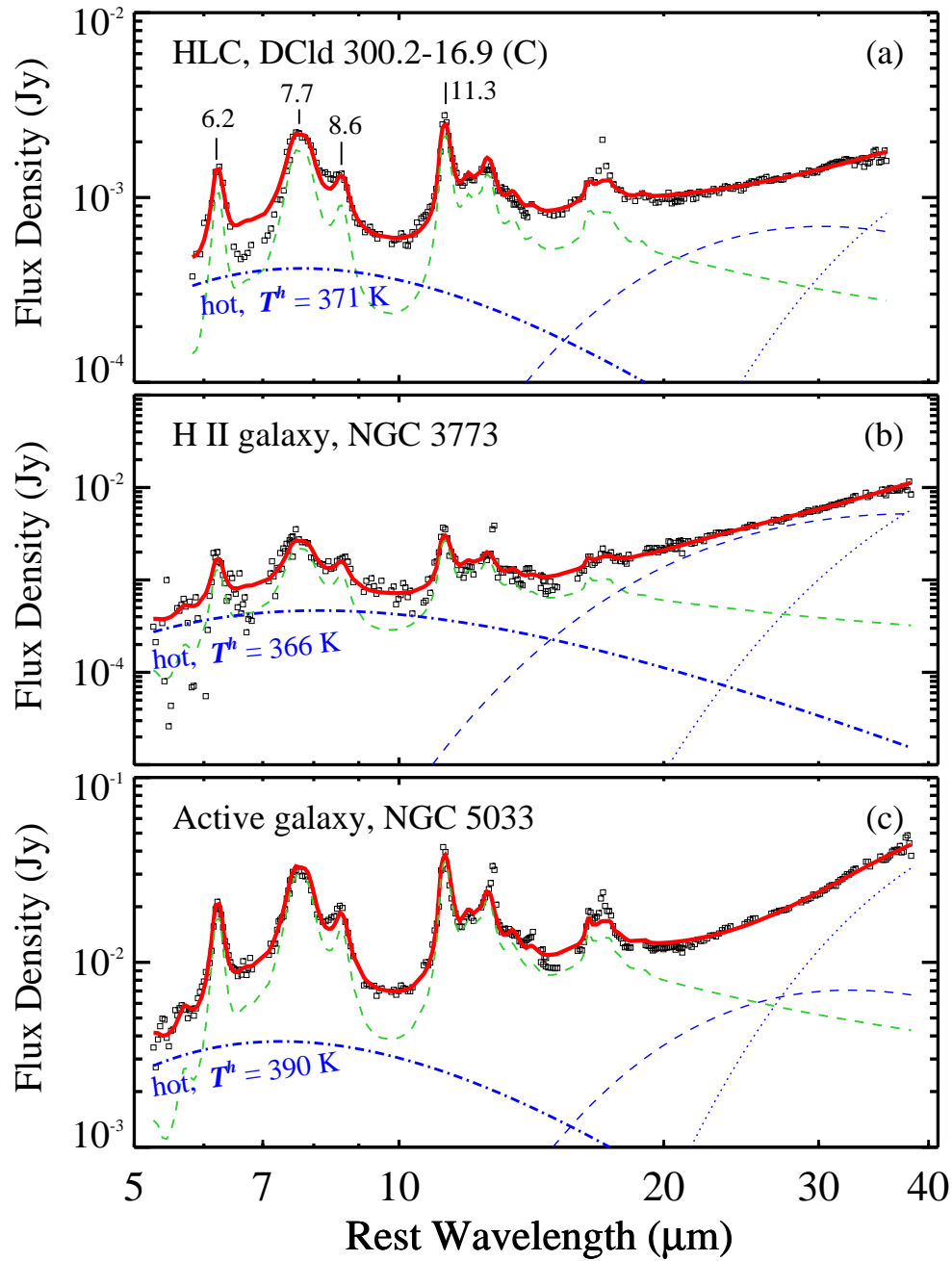}}
\caption{\footnotesize
Spectral decomposition of the MIR spectra of ({\it a}) HLC DCld~300.2-16.9 (C), 
({\it b}) the H~II galaxy NGC~3773, and ({\it c}) the Seyfert galaxy NGC~5033.
Each panel shows the observed {\it Spitzer}/IRS spectrum (black squares) and
the best-fit model (solid red line), comprising a theoretical PAH template 
(green dashed line; main PAH features labeled in top panel) and three dust 
components (hot: blue dot-dashed line; warm: blue dashed line; cold: blue 
dotted line).  The temperature of the hot component is labeled.  
The main PAH emission features at 6.2, 7.7, 8.6 and 11.3$\mum$ are also labeled 
in the top panel. 
\label{fig:show_fit} }
\end{center}
\end{figure}

\begin{figure}
\begin{center}
\resizebox{0.8\hsize}{!}
{\includegraphics{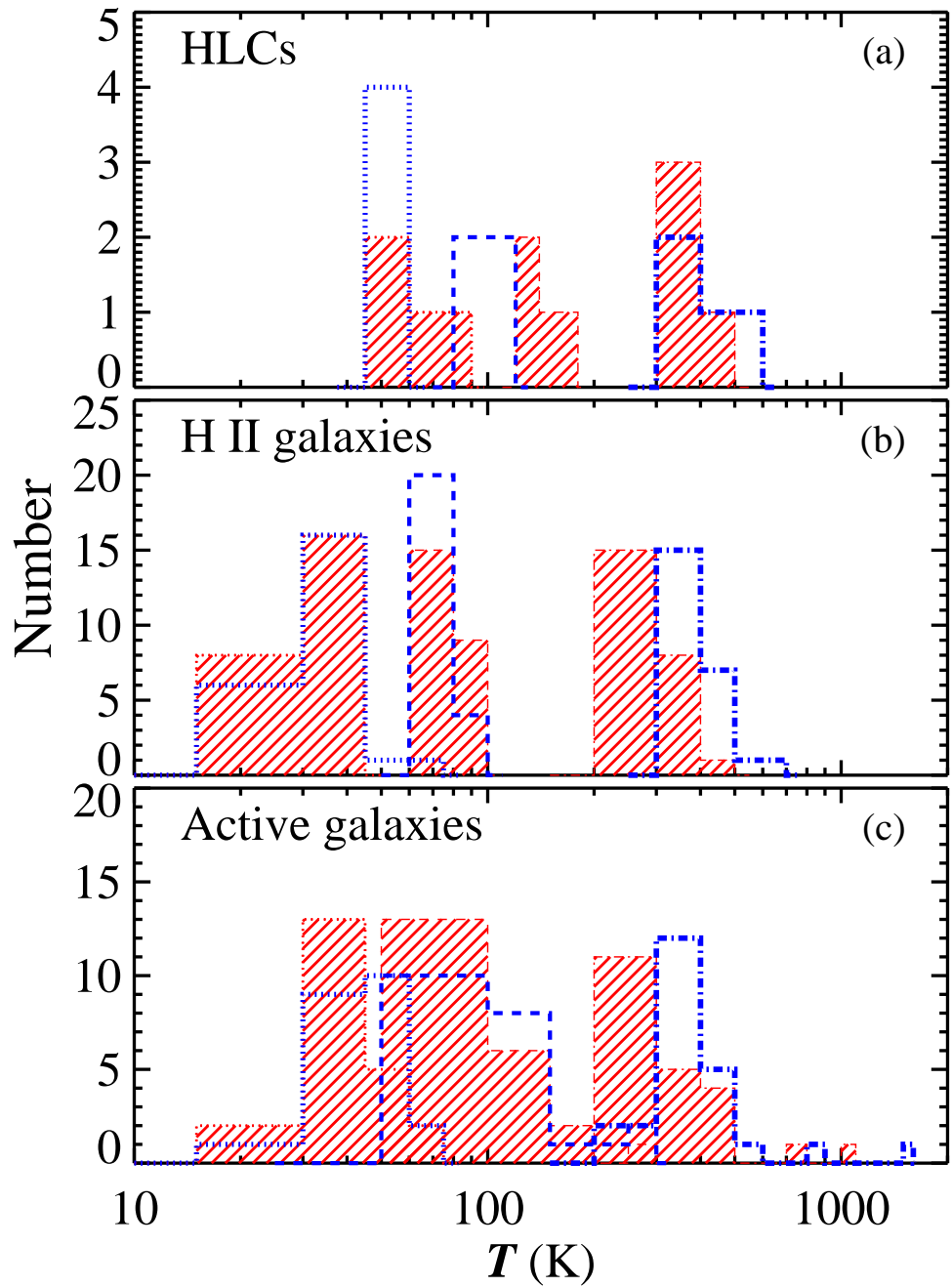}}
\caption{\footnotesize
Temperature distributions of the three dust components. The one   
derived based on theoretical PAH template are plotted in blue color 
(hot: dot-dashed line; warm: dashed line; cold: dotted line) for ({\it a}) HLCs, 
({\it b}) H II galaxies, and ({\it c}) active (LINER and Seyfert) galaxies.
The one derived based on an alternate empirical PAH template 
(CAFE; Marshall \etal 2007) are in red shaded area wrapped with 
different lines (hot: dot-dashed line; warm: dashed line; cold: dotted line) 
for ({\it a}) HLCs, ({\it b}) H II galaxies, and ({\it c}) active (LINER and Seyfert) galaxies. 
\label{fig:temp_hist} }
\end{center}
\end{figure}

In Figure~\ref{fig:show_fit}, we illustrate the decomposition of the
mid-IR spectra of a Galactic high-latitude cloud (HLC), a normal
star-forming galaxy, and a low-luminosity active galaxy. Both classes
of objects under investigation---Galactic HLCs and nearby
galaxies---exhibit a prominent mid-IR continuum consistent with
thermal radiation from very hot dust grains. The temperatures span 
a relatively narrow range, from $\simali$300 to $\simali$400$\K$ 
(Figure~\ref{fig:temp_hist}, blue histogram).  Not only are the
temperatures high, the most surprising aspect is that they are
relatively uniform across extremely diverse environments, from
Galactic HLCs to local star-forming galaxies and active galaxies of
different types, from LINERs to Seyferts, which cover a range of
ionization levels and black hole accretion rates (Ho 2008, 2009).

The derived temperatures are very robust against the details of the
spectral decomposition. Our companion paper (Xie \etal 2018) shows 
that for grain sizes $a\simlt 20\Angstrom$ or $a \simlt 50\Angstrom$, 
the PAH spectrum for $\lambda$ \lax\ 15$\mum$ is quite uniform 
for a wide range of photon densities from $U=1$ to $U=10^{6}$ 
(see Figure~2 in Xie et al.\ 2018), and that
the derived PAH and dust temperatures also vary little among fits
adopting PAH templates with $U\simlt10^{4}$ 
(see Figure~12 in Xie et al.\ 2018). 
To ascertain the robustness
of the derived dust temperatures, we also consider an alternate
empirical PAH template derived from the ``Continuum And Feature
Extraction'' (CAFE) method of Marshall \etal (2007) and repeat the fits.
We find that the derived dust temperature distributions remain
virtually unchanged relative to those derived from our theoretical PAH
template (see Figure~\ref{fig:temp_hist}). Also, the derived hot dust
temperature is relatively insensitive to the omission of longer
wavelength data extending to the far-IR (see Xie et al.\ 2018). Given
the pervasiveness of the hot-dust component, why has it gone largely
unnoticed?  The reason, we suspect, stems from the sheer complexity of
the mid-IR spectra, especially the overwhelming dominance of the many
strong, overlapping PAH features, whose broad and extended line
profiles effectively mask the local continuum.  Absent an effective,
physically well-motivated method to decouple  the PAH spectrum, the
underlying mid-IR continuum becomes difficult to disentangle. In
\S\ref{sec:size}, we discuss in more detail the difference
between PAHs and the emitter of the hot continuum derived here.

The equilibrium temperature of dust depends on 
the strength of the radiation field to which it 
is exposed, as $T\propto U^{1/(4\,+\,\beta)}$, 
where $U$ describes the starlight intensity 
that heats the grains 
and $\beta$ is the power-law index of 
dust emissivity in the far-IR, 
assumed $\lambda^{-\beta}$, 
with $\beta\approx1.5-2$ for greybody emission. 
The above scaling relation indicates that an increase 
of starlight intensity by a factor of 100 only induces a 
temperature rise of a factor of $\simali$2.  
Accumulating evidence shows that the dust temperature 
in the ISM, both near and far, 
occupies a very narrow range.  
For instance, the emission of 
the diffuse interstellar dust of the Milky Way, 
with an average star formation rate of 
$\simali$1.7$\Msun\yr^{-1}$ (Licquia \& Newman 2015), 
peaks at $\simali$18$\K$ (Planck Collaboration 2011). 
Extensive {\it Herschel} surveys of local star-forming 
and starburst galaxies reveal that the bulk of 
the ISM mass has temperatures $T$\,$\approx$15--30$\K$ 
(Dunne \etal 2011; Auld \etal 2013).  
High-redshift ($z$\,$\approx$\,1.4--3) submillimeter galaxies 
also display a narrow range of dust temperatures 
(Scoville \etal 2014).  The narrow temperature distribution 
for the bulk of the ISM, at both low and high redshifts, 
indicates that the interstellar radiation field of 
normal star-forming or starburst galaxies does not 
vary significantly.  Indeed, the IR spectral energy 
distribution (SED) of the SINGS galaxies all peak 
around 100$\mum$, corresponding to dust temperature 
of $\simali$20--30$\K$ (Dale \etal 2012).

If the observed hot dust emission described 
in this work is ascribed to an equilibrium 
temperature, it would require a dust-heating 
starlight intensity a few million times higher 
than that found in typical star-forming galaxies.  
In principle, this can be achieved from intense 
starburst activity confined to a sufficiently 
compact region.  However, such a scenario is 
utterly implausible for HLCs, which have no 
ongoing embedded star formation and are located 
far from the Galactic disk ($70\pm15\pc$).  
If the observed mid-IR emission of 
the HLC DCld 300.2$-$16.9 is from equilibrium 
dust radiation from external heating, 
there should be a temperature gradient 
across the cloud varying with distance 
from the Galactic disk.  No such trend is seen.    
As for the H~II galaxies considered here, 
they have modest integrated star formation rates of 
$\simlt$\,1$\Msun\yr^{-1}$ to $\simali$15$\Msun\yr^{-1}$, 
distributed broadly across the entire disk.  
The IRS slit only captures a portion of 
the star formation from the central kpc,
which does not contribute a significant 
amount of heating photons. The same can be said 
of the active galaxies in our sample.  
While technically classified as accretion-powered sources, 
the level of nuclear activity in these nearby LINERs 
and Seyferts is incredibly feeble (Ho 2008, 2009). 
It is impossible for these low-luminosity AGNs 
to provide the radiation intensity to 
heat the dust grains to the requisite high temperatures.  
Furthermore, if the observed high temperature is 
generated from hot, optically thin dust heated by 
intense UV photons, we would expect to see silicate 
emission at $\simali$10 and 18$\mum$.  
None of the objects in our sample 
shows silicate emission; some exhibit, 
at most, mild silicate absorption.  
Therefore, equilibrium heating can be ruled out 
as the emission mechanism for the strikingly 
high temperatures observed in the mid-IR continuum.  

We suggest that the ubiquitous presence of 
high-temperature dust originates from nano grains 
heated by single-photon absorption 
(i.e., stochastic heating), in the manner 
first proposed 35 years ago (Sellgren \etal 1983, 
Sellgren 1984; Weiland \etal 1986). 
The temperature attained through stochastic heating scales only with
the grain size and the average photon energy and is independent of the
distance from the heating source.  Such a mechanism provides a natural
explanation for the relatively invariant mid-IR continuum emission
detected toward sources as diverse as Galactic HLCs and low-luminosity
active galaxies. 
The grain size corresponding to this emission is 
$\simlt$\,1$\nm$ (see \S~\ref{sec:size}).

\subsection{Grain Size\label{sec:size}}
Upon absorption of a photon of energy $h\nu$, a grain undergoing
single-photon heating will be very quickly ($<10^{-3}\s$) heated to
its peak temperature $\Tpeak$ and then rapidly cools down in a few
seconds by emiting most of the absorbed energy at $\Tpeak$ (see Draine
\& Li 2001). The peak temperature is determined from 
\begin{equation}
\label{eq:TpeakC}
{\int_{0}^{\Tpeak} C(T)\,dT = h\nu}~,
\end{equation}
\noindent
where $C(T)$ is the specific heat of the grain at temperature $T$, $h$
is the Planck constant, and $\nu$ is the frequency of the absorbed
photon. Let $\TD$ be the Debye temperature of the grain material (\eg
$\TD\approx420\K$ for graphite, $\TD\approx542\K$ for quartz SiO$_2$;
see Draine \& Li 2001) and $\Natom\propto a^3$ be the number of atoms in 
a grain of size $a$. For $T\ll\TD$, 
\begin{equation}
\label{eq:specheat}
\frac{C(T)}{\Natom} = \frac {12k\pi^{4}}{5} 
     \left(\frac{T}{\TD}\right)^3 ~~,
\end{equation}
\noindent
and therefore,
\begin{equation}
\label{eq:Tpeakhnu}
\Tpeak = \left(\frac{5}{3k\pi^4}\right)^{1/4}
         \left(h\nu\right)^{1/4}
         \Natom^{-1/4}\,\TD^{3/4} 
         \propto a^{-3/4}\,\left(h\nu\right)^{1/4}~~.
\end{equation}
\noindent
For $T\gg \TD$, $C(T)\approx 3\Natom\,k$ (i.e., the Dulong-Petit law), 
where $k$ is the Boltzman constant, and therefore $\Tpeak$ will be 
appreciably higher than that given by eq.\,\ref{eq:Tpeakhnu}, 
but it is always true that $\Tpeak\propto a^{-3/4}\,\left(h\nu\right)^{1/4}$.

We can estimate the size of the ultrasmall grains responsible for the
hot ($T^h\approx300-400\K$) component from eq.\,\ref{eq:TpeakC}. We
take $\Tpeak\approx T^h$ because grains heated by single photons
radiate most of their energy at $\Tpeak$ (Draine \& Li 2001). Since
$\Tpeak$ is comparable to $\TD$, we cannot simply use
eq.\,\ref{eq:specheat} for $C(T)$ which is valid for $T\ll \TD$.
Following Draine \& Li (2001), we will approximate the specific
heats of graphite and silicate materials in terms of the following 
Debye models\footnote{There is a typographical error
in the expression of $f_n(x)$ in eq.\,10
of Draine \& Li (2001).
}:
\beq
C_{\rm gra}(T)= (\Natom-2) k
\left[
f_2^\prime\left(\frac{T}{863\K}\right) 
+ 2\,f_2^\prime\left(\frac{T}{2504\K}\right)
\right]~~,
\label{eq:graph_spec_heat}
\eeq
\beq
C_{\rm sil}(T)= (\Natom-2)k
\left[
2f_2^\prime\left(\frac{T}{500\K}\right)
+f_3^\prime\left(\frac{T}{1500\K}\right)
\right]~~,
\label{eq:sil_spec_heat}
\eeq
\beq
f_n(x) \equiv n\int_0^1 \frac{y^n dy}{\exp(y/x)-1} ~~~,~~~
f_n^\prime(x)\equiv \frac{d}{dx}f_n(x) ~~~.
\label{eq:f_n}
\eeq

For the photon energy $h\nu$,
we consider the {\it mean} photon energy $\langle h\nu\rangle$ 
that an ultrasmall grain would absorb when exposed to 
starlight of radiation density $u_\nu$ (or radiation intensity 
$cu_\nu/4\pi$, where $c$ is the speed of light):
\begin{equation}
\label{eq:num7}
\langle h\nu\rangle = 
    \frac{\int \Cabs(a,\nu)\,cu_\nu\,d\nu}
    {\int \Cabs(a,\nu)\,cu_\nu/h\nu\,d\nu} ~~,
\end{equation}
\noindent
where $\Cabs(a,\nu)$ is the absorption cross section of 
the grain of size $a$ at frequency $\nu$. If we approximate 
the radiation field of normal star-forming galaxies by the 
MMP83 ISRF of the solar neighborhood (Mathis \etal 1983), 
we obtain $\langle h\nu \rangle \approx7.5\eV$ for 
nanometer-size silicate grains and $\langle h\nu \rangle \approx3.5\eV$ 
for nanometer-size graphitic grains. Similarly, for starburst galaxies, 
if we approximate their radiation field by the UV, visible, 
and near-IR spectrum of NGC\,7714 (Brown \etal 2014), 
we estimate $\langle h\nu \rangle \approx8.8\eV$ 
for nanometer-size silicate grains and 
$\langle h\nu \rangle \approx4.8\eV$ for nanometer-size 
graphitic grains. The mean photon energy absorbed by 
nanometer-size grains is insensitive to grain size because 
such small grains, even in the UV, are in the Rayleigh regime 
and thus $\Cabs(a,\nu)/a^3$ is independent of $a$.

With $\langle h\nu \rangle \approx7.5\eV$ and 8.8$\eV$, 
we derive, respectively, $a\approx8.0\pm0.6\Angstrom$ 
and $8.5\pm0.7\Angstrom$ for silicate grains of 
$\Tpeak\approx300-400\K$. Similarly, with 
$\langle h\nu \rangle \approx3.5\eV$ and 4.8$\eV$, 
we derive $a\approx7.7\pm0.8\Angstrom$ and 
$8.4\pm0.9\Angstrom$ for graphitic grains of 
$\Tpeak\approx300-400\K$. 
We will call these ultrasmall grains 
(which are responsible for 
the mid-IR continuum at $\lambda\simlt10\mum$) 
``nanometer-size'' grains.
These nano grains differ from 
the so-called ``very small grains'' (VSGs)
reported by Sellgren \etal (1983) and Sturm \etal (2000) 
that emit at $\simali$3--5$\mum$.  The latter
are smaller in size,
as they should be transiently heated to 
a higher peak temperature 
(i.e., $\Tpeak$\,$\gtsim$\,800$\K$).
These nano grains are also distinct from 
the VSG component invoked by grain models 
in the post-{\it IRAS}\ era (e.g., see D\'esert et al.\ 1990; 
Siebenmorgen \& Kruegel 1992; Dwek \etal 1997; 
Li \& Greenberg 1997; Zubko et al.\ 2004;  
Compiegne \etal 2011). 
The nanometer-size component inferred here 
emits mostly at $\lambda$\,$\simlt$\,10$\mum$ 
and negligibly at $\lambda$\,$\simgt$\,20$\mum$, 
while the VSGs invoked by the aforementioned models
are much larger in size and dominate 
the {\it IRAS} 25$\mum$ emission 
and in part the {\it IRAS} 60$\mum$ emission. 
Also, the nano component discussed here differs 
from the VSG component of Boissel \etal (2001)
and Rapacioli \etal (2005) derived from 
the singular value decomposition method 
coupled with a Monte Carlo search algorithm.
The latter is essentially a PAH cluster and exhibits 
a broad band at $\simali$7.8$\mum$,
while the nano component discussed here
emits a smooth continuum.

The nanometer-size grains inferred here 
are also different from PAHs.
The so-called ``unidentified IR emission'' (UIE) bands
at 3.3, 6.2, 7.7, 8.6, and 11.3$\mum$ commonly seen in 
various astronomical environments have been attributed 
to tiny carbonaceous grains or very large molecules of 
size $a < 50\Angstrom$ and carbon atoms 
$N_{\rm C} < 2 \times10^{4}$ 
(\eg see Figure~7 of Draine \& Li 2007), 
with PAH molecules as the mostly likely carrier 
(L\'eger \& Puget 1984; Allamandola et al.\ 1985). 
PAHs emit remarkably similarly in objects spanning
several orders of magnitude in radiation density 
(\eg Boulanger \etal 2000). 
Xie \etal (2018) show that PAH spectra 
barely change in environments as diverse
as Galactic HLCs, nearby star-forming 
and starburst galaxies, and even in galaxies 
hosting AGNs of both low and high luminosity. 
The grains (or large molecules) contributing 
to the PAH emission features have similar nature 
to grains responsible for 
the uniform hot continuum emission: 
both are excited by stochastic heating from 
single-photon absorption (Draine \& Li 2001, 2007).
However, experimentally, gas-phase, 
free-flying PAH molecules of several tens 
of carbon atoms do not show strong 
continuum opacity at $\simali$1--5$\mum$ 
(see Bauschlicher et al.\ 2018 and references therein).
Also, An \& Sellgren (2003) found that in NGC\,2023, 
a prototypical reflection nebula, the 3.3$\mum$ PAH 
emission and the underlying continuum emission 
at $\simali$2$\mum$ are spatially separated, 
suggesting that PAHs are not the carrier of 
the 2$\mum$ continuum emission.\footnote{%
   Kahanp{\"a}{\"a} \etal (2003) correlated 
   the total PAH emission ($I_{\mathrm {PAH}}$) 
   at $\simali$5.8--11.6$\mum$ 
   with the {\it IRAS} 12 and 100$\mum$ 
   surface brightness 
   ($I_{\mathrm {12}}, I_{\mathrm{100}}$) 
   measured across the Galaxy. They found that $I_{\mathrm
  {PAH}}$ correlates with $I_{\mathrm {12}}$ and $I_{\mathrm {100}}$
  fairly well. While a close correlation between $I_{\mathrm {PAH}}$ and
  $I_{\mathrm {12}}$ is expected since the {\it IRAS} 12$\mum$ emission
  is predominantly contributed by PAHs (\eg see Figure~8 of Li \& Draine 2001),
  the correlation between $I_{\mathrm {PAH}}$ 
  and $I_{\mathrm {100}}$ does not necessarily 
  mean that PAHs and the grains responsible 
  for the 100$\mum$ continuum emission 
  are physically associated. The latter correlation 
  could merely reflect the fact that both PAHs 
  and the submicron-size grains --- the dominant 
  emitter of the 100$\mum$ thermal continuum 
  --- correlate with the total amount of interstellar 
  material (\eg the hydrogen column density 
  $N_{\mathrm {H}}$). In order to properly 
  examine whether PAHs and the carriers 
  of the 100$\mum$ emission are indeed 
  physically related,  one needs to eliminate 
  their common correlations with $N_{\mathrm {H}}$, 
  by comparing $I_{\mathrm {PAH}}/N_{\mathrm H}$ 
  with $I_{\mathrm {100}}/N_{\mathrm H}$.
  }
Alternatively, Kwok \& Zhang (2011) 
suggest that organic nanoparticles 
with a mixed aromatic-aliphatic structure 
could be responsible for 
\textit{both} the PAH emission bands 
\textit{and} the underlying continuum. 
Future spatially resolved observations 
(\eg \textit{JWST}) that can map the two components 
will provide new insight into their physical separations 
and chemical carriers.

\subsection{Chemical Carrier\label{sec:composition}}
\begin{figure}  
\begin{center}
\resizebox{0.6\vsize}{!}
{\includegraphics{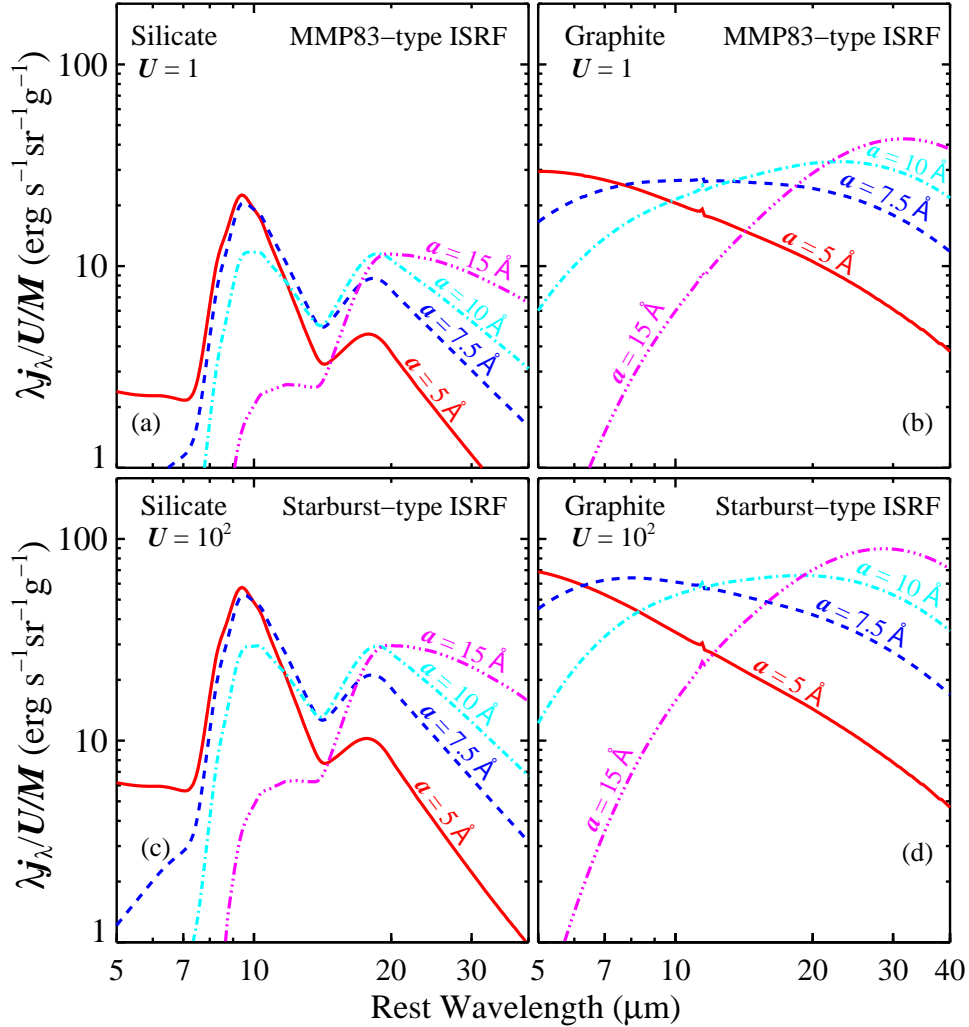}} 
\caption{%
   IR emissivity (normalized by dust mass $M$ 
   and starlight intensity $U$)
   for single-size silicate (a, c) 
   and graphite (b, d) grains of 
   $a = 5\Angstrom$ (red solid line), 
   $a = 7.5\Angstrom$ (blue dashed line), 
   $a = 10\Angstrom$ (cyan dot-dashed line),
   and $a = 15\Angstrom$ (magenta dot-dot-dashed line)
   illuminated by the MMP83-type ISRF of $U=1$ (a, b)
   and the starburst-type ISRF of $U=100$ (c, d).
   \label{fig:nano_irem}
   }
\end{center}
\end{figure}

Using the technique of Draine \& Li (2001) developed for modeling 
the vibrational excitation of stochastically heated ultrasmall grains,
we calculate the IR emissivity of spherical nanometer-size grains of 
different chemical compositions of radii $a\,=\,5, 7.5, 10$, and 
$15\Angstrom$, respectively. 
We consider silicate and graphite grains. 
We consider a heating source similar to that 
of the MMP83 solar neighborhood ISRF 
(``MMP83-type ISRF''; Mathis et al.\ 1983) 
and a harder, more intense starburst ISRF 
(approximated by the UV, visible and near-IR spectrum
of NGC\,7714; Brown et al.\ 2014) 
with $U\,=\,100$  
(``starburst-type ISRF'').
As illustrated  in Figure~\ref{fig:nano_irem}, 
nanometer-size silicate grains emit prominent features at 9.7$\mum$
(Si--O) and 18$\mum$ (O--Si--O), whether the ISRF is MMP83-type or
starburst-type (see Figure~\ref{fig:nano_irem}a, c).  These are not seen 
in any of the sources discussed in this work. Therefore, we can exclude 
nanometer-size amorphous silicates as the main carrier of the hot dust 
emission observed in the mid-IR.
In contrast, the free electrons of nanometer-size graphitic grains
emit a smooth continuum in the mid-IR 
(see Figure~\ref{fig:nano_irem}b, d).
The lack of any prominent spectral features
from nano graphite is consistent with 
the ``derived'' featureless hot component emission.

At a first glance, 
nanometer-size amorphous carbon (AC) grains 
could also be a plausible carrier for 
the hot component, for they also emit 
a featureless continuum in the mid-IR.
In the diffuse ISM, however, AC grains are likely hydrogenated, as
revealed by the 3.4$\mum$ aliphatic C--H stretch {\it absorption} band
(Pendleton \& Allamandola 2002). If nanometer-size AC grains are
present in the ISM, e.g., resulting from the collisional fragmentation
of bulk, submicron-size hydrogenated amorphous carbon (HAC) grains,
unless they are completely dehydrogenated, they would emit at
3.4$\mum$ because of single-photon heating. Observationally, the
3.4$\mum$ aliphatic C--H {\it emission} feature is mostly seen in
cool, UV-poor reflection nebulae and proto-planetary nebulae and is
often much weaker than the 3.3$\mum$ aromatic C--H band (Yang \etal
2013). Therefore, it is less likely that the nanometer-size component
inferred here is from nanometer-size AC grains unless their H atoms
are completely knocked off. 

Like graphite, nanometer-size iron grains also 
emit a featureless continuum in the IR. 
While they seem to be stable against sublimation
in the ISM (Hensley \& Draine 2017b) and 
have recently been invoked by to account for
the ``anomalous microwave emission''
(Hensley \& Draine 2017a),
it is not clear if they could survive 
against oxidation (Jones 1990). 
Nanometer-size iron oxides would 
exhibit several spectral features, 
e.g., FeO at $\simali$21$\mum$,
Fe$_2$O$_3$ at $\simali$20.5$\mum$ and 27.5$\mum$, 
and Fe$_3$O$_4$ at $\simali$16.5$\mum$ and 24$\mum$ 
(Henning \& Mutschke 1997; Koike \etal 2017). 
However, these features are not seen in the ISM.

In principle, the spectral profile of the hot component should 
allow us to constrain the composition of the nanometer-size grain population.  
We do not do so quantitatively because the approach adopted here to determine 
the spectral shape of the hot component was simplified (see eq.\,\ref{eq:num1}), 
and the actual emission spectrum of a nanometer-size grain is considerably wider 
than that approximated by a single temperature (Draine \& Li 2001). 
Therefore, it would be an over-interpretation to employ the emission spectrum 
of the hot component to infer the nanometer-size grain composition (and size distribution).  
Nevertheless, we propose that nano graphite is primarily responsible for 
this radiation component, as nanometer-size graphitic grains emit a smooth continuum 
in the mid-IR caused by the free electrons of graphite, a semi-metal, 
and carbon is an abundant element, much more abundant than iron. 
Presolar graphite grains have been identified in primitive meteorites
through isotope anomaly. Nanometer-size graphite grains are a natural
product of shock-induced shattering of large, submicron-size graphite
grains, or thermal processing of nanodiamonds (Andersson et al.\ 1998) 
which are also abundant in primitive meteorites (Lewis et al.\ 1987). 
Presolar nanodiamonds are abundant in primitive meteorites 
and the detection of nanodiamonds through their characteristic 
emission features at 3.43 and 3.53$\mum$ have been reported 
in HD\,97048 and Elias\,1, two Herbig Ae/Be stars,
and in HR\,4049, a post-asymptotic giant branch star 
(see Li 2004a and references therein).
However, nanodiamonds are unlikely responsible 
for the hot continuum emission reported in this work
since neither the 3.43$\mum$ emission feature 
nor the 3.53$\mum$ emisison feature 
expected from nanodiamonds 
is seen in the ISM of the Milky Way 
and external galaxies.

\subsection{Mass Fraction\label{sec:mass}} 

Let $f_{\mathrm {mass}}({\rm nano})$ be the fraction of the total dust
mass in the nanometer-size dust component, $f_{\mathrm {IR}}({\rm
  nano})$ be the corresponding fraction of the total IR power,
$\kabsnano(\nu)$ be the corresponding mass absorption coefficient, and
$\langle \kappa_{\mathrm {abs}}(\nu)\rangle$ be the mean mass
absorption coefficient of all dust grains in a galaxy. For energy
balance between absorption and emission, we obtain
\begin{equation}
\label{eq:num8}
f_{\mathrm {mass}}({\rm nano}) = f_{\mathrm {IR}}({\rm nano}) 
  \frac{\int \langle \kappa_{\rm abs}(\nu)\rangle\,cu_\nu\,d\nu}
  {\int \kabsnano(\nu)\,cu_\nu\,d\nu} ~~.
\end{equation}
\noindent
This does not involve any dust temperature which is known to be a major source of uncertainties in estimating dust mass. We assume that the bulk dust mixture of the galaxies considered here is more or less the same as that of the Milky Way galaxy (Li \& Draine 2001). For the nanometer-size component, we calculate $\kabsnano(a,\nu)\equiv \Cabs(a,\nu)/M$ from graphite grains of $a=10\Angstrom$, where $M=\left(4\pi/3\right) a^3 \rho$ is the mass of a graphite grain of size $a$ and $\rho=2.24\g\cm^{-3}$ is the mass density of graphite. Since nanometer-size grains are in the Rayleigh regime from the UV to the far-IR, $\kabsnano(a,\nu)$ is independent of $a$.

Estimating $f_{\mathrm {IR}}({\rm nano})$ requires complete coverage of the SED from the near-IR to the far-IR. While sufficient wavelength coverage exists for many of the objects analyzed in this study, the SINGS galaxies are angularly large, and aperture mismatch between the {\it Spitzer}\ IRS and {\it Herschel}\ observations is a serious problem.  Thus, for this portion of the analysis we make use of archival observations from the Great Observatories All-Sky LIRG Survey (GOALS, Armus \etal 2009), a sample of 201 nearby ($z\approx 0.09$) IR-luminous galaxies with complete mid-IR spectra from \textit{Spitzer}/IRS and photometry spanning $\simali$12--500$\mum$ from {\it IRAS}\ and {\it Herschel}.  We proceed in a stepwise fashion to select a subset of the GOALS objects for which we can compile SEDs that properly sample the same physical scale, despite the vastly different resolutions between the mid-IR and far-IR.
We begin by scaling the flux density of the IRS spectra from the short-low module ($\simali$5.2--14.5$\mum$; slit width $3.6^{\prime\prime}$) to match the spectra from the long-low module ($\simali$14.0--38$\mum$; slit width $10.5^{\prime\prime}$). We only choose galaxies with sufficiently large distances such that their optical diameters fit within the dimensions of the long-low aperture.  As the point-spread function of all the {\it IRAS}\ bands comfortably covers the total extent of these galaxies, we require that the IRS 25$\mum$ flux density matches the {\it IRAS} 25$\mum$ photometry (Sanders \etal 2003), allowing for $\simali$10\%--15\% variation to account for systematic uncertainties between the two datasets.  Next, we verified that the {\it IRAS}\ 60 and 100$\mum$ photometry closely matches the {\it Herschel}\ measurements at 70 and 100$\mum$ (Chu \etal 2017).  This ensures that the {\it Herschel}\ photometry out to 500$\mum$ can be used.  Of the 60 galaxies that match these criteria, we finally select 28 that are starburst-dominated systems according to classifications in the NASA/IPAC Extragalactic Database\footnote{\tt http://ned.ipac.caltech.edu}, the majority of which are based on mid-IR diagnostics (Armus \etal 2007). The IRS spectra of all 28 galaxies exhibit unambiguously strong PAH emission.   In Table~\ref{tab:goals}, we present the basic parameters and {\it Herschel}\ photometry for this subset of GOALs galaxies.

\begin{figure}  
\begin{center}
\resizebox{0.6\vsize}{!}
{\includegraphics{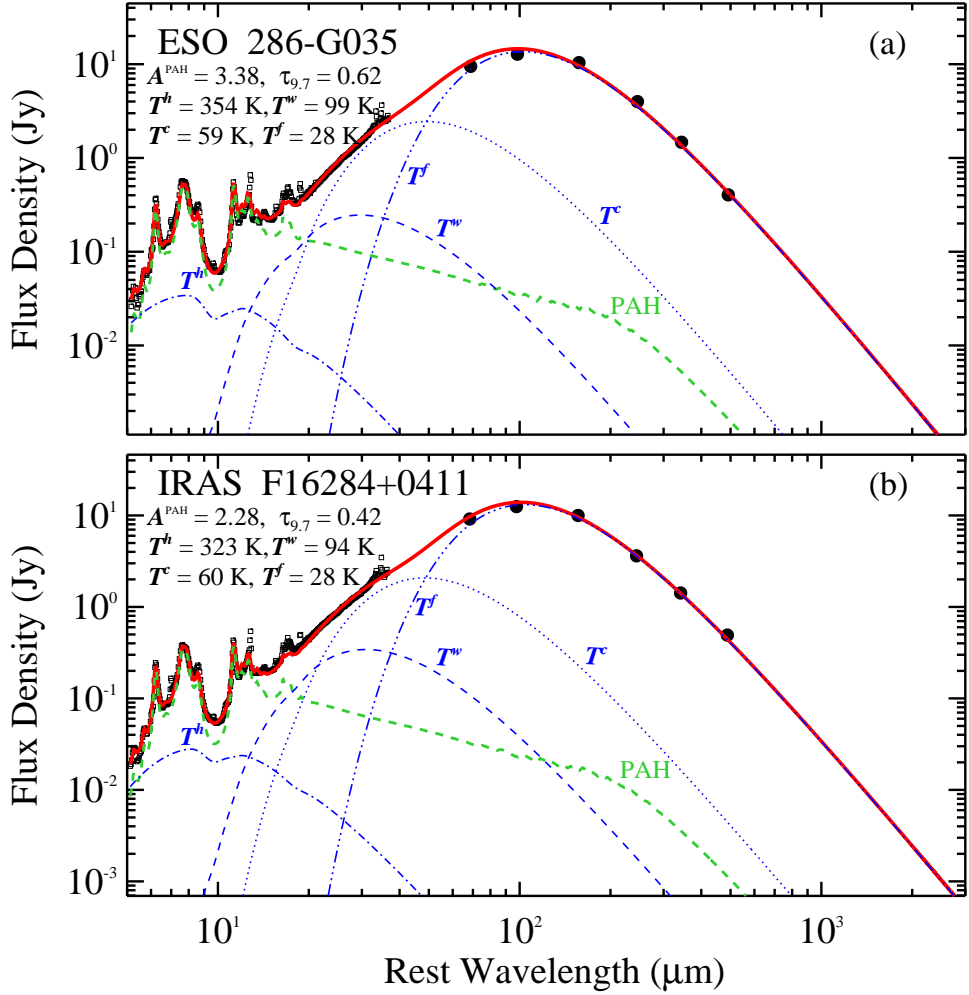}} 
\caption{
Decomposition of the SED of (a) ESO\,286-G035 and (b) IRAS\,F16284+0411.  The SED is extended to the far-IR  using photometric measurements from {\it Herschel}\ at 70, 100, 160, 250, 350, and 500$\mum$.  Each panel shows the observed {\it Spitzer}/IRS spectrum (black squares), far-IR photometry (black filled circles), and the best-fit model (solid red line), comprising a theoretical PAH template (green dashed line) and four dust components (hot: blue dot-dashed line; warm: blue dashed line; cold: blue dotted line; extreme cold: blue dot-dot-dot-dashed line. 
\label{fig:mir_fir_sed_fit}}
\end{center}
\end{figure}

\begin{figure}  
\begin{center}
\resizebox{0.6\vsize}{!}
{\includegraphics{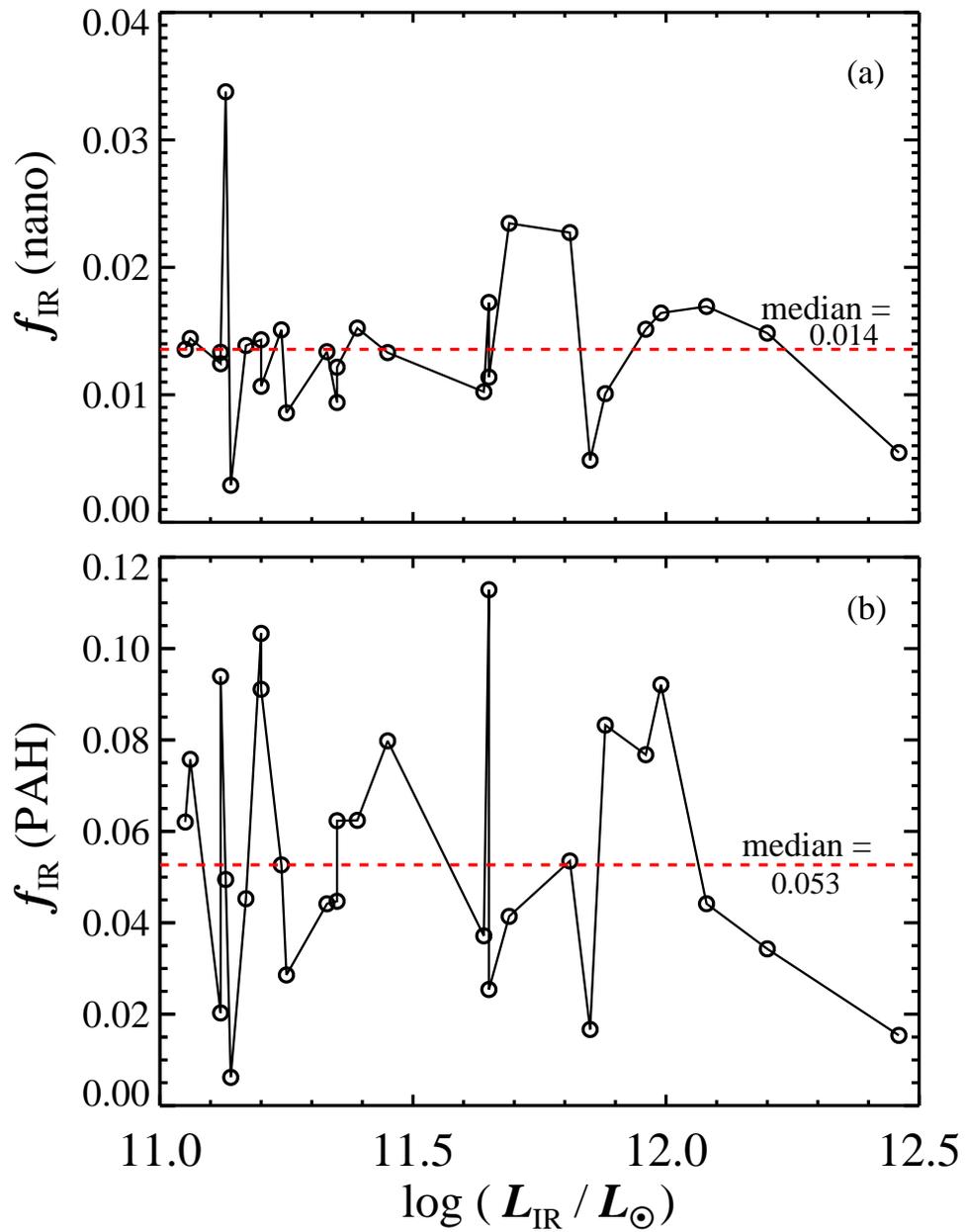}} 
\caption{%
The fraction of the total IR power emitted by (a) the nano dust component 
and (b) PAHs.
\label{fig:nano_pah_frac}
}
\end{center}
\end{figure}

We fit the complete IR SED using the methodology described in our companion paper Xie \etal (2018), adopting a PAH template for the PAH emission and four modified blackbodies to account for the dust continuum emission (Figure~\ref{fig:mir_fir_sed_fit}). We derive $f_{\mathrm {IR}}({\rm nano})$, the flux fraction radiated from nanometer-size dust, by dividing the integrated hot dust emission with the total radiation from 5 to 3000$\mum$.  The results are shown in Figure~\ref{fig:nano_pah_frac}a, where we also plot the flux fraction radiated from PAHs (see Figure~\ref{fig:nano_pah_frac}b). The nanometer-size dust component accounts for $f_{\mathrm {IR}}({\rm nano})\approx 1.4\%$, independent of the total IR luminosity of the galaxies.  PAHs behave similarly, with a median IR power fraction of $\simali$5\%.  An $f_{\mathrm {IR}}({\rm nano})$ of 1.4\% translates into $f_{\mathrm {mass}}({\rm nano})\approx 0.4\%$, for a silicate-graphite mixture like that of the Milky Way diffuse ISM (Weingartner \& Draine 2001). This corresponds to a mass fraction of $\simali$1.5\% of the total carbon dust, lower than that of PAHs by a factor of $\simali$3. 

We note that, in order to account for the $\simali$3--5$\mum$ 
continuum emission detected by Sellgren et al.\ (1983)
and Sturm et al.\ (2000),
the astro-PAH model 
(Li \& Draine 2001, Draine \& Li 2007)
from which the theoretical PAH template 
was calculated and adopted here 
incorporates a small amount of ``continuum'' opacity, 
equal to $\simali$1\% of bulk graphite, 
although the exact nature of 
this continuum opacity is unclear. 
If free-flying PAHs do not have 
any continuum opacity in the near IR, 
the fractional emission contributed by 
the hot component would be even higher 
and therefore the quantity of
the nanometer-size dust would be larger, 
and our conclusion of the presence of 
nanometer-size dust would be even more robust. 

The 2175$\Angstrom$ interstellar extinction bump is often attributed to small graphite grains
(\eg see Stecher \& Donn 1965; Draine 1988; Mishra \& Li 2015, 2017). 
With such a small mass fraction, the nanometer-size graphite component inferred here from the hot continuum emission at $\lambda$\,$\simlt$10$\mum$ cannot be a major contributor to the 2175$\Angstrom$ extinction bump since the latter requires its carrier to lock up $\simali$10--15\% of the total interstellar carbon abundance (Draine 1989).

Lastly, we note that we have not considered high-luminosity active galactic nuclei (\eg quasars).  The mid-IR continuum emission from nanometer-size dust, if present, would be difficult to isolate from the hot dust emission from the nucleus-heated dusty torus (Zhuang \etal 2018).  Moreover, stochastic heating is unlikely to be effective in light of the high average photon energy in these luminous systems. As the photon absorption timescale becomes short and comparable to the cooling time, small grains cannot completely cool off in between photon absorption events and hence will eventually attain temperature equilibrium. Large grains can also reach very high temperatures from the frequent absorption of heating photons and peak in the mid-IR (Xie \etal 2017).

\section{Summary}
We found a hot dust continuum emission 
at $\lambda$\,$\simlt$\,10$\mum$ of
temperature $\simali$300--400$\K$ that is
widespread in different 
astronomical environments. 
This hot continuum component, 
closely coupled to the PAH emission features, 
is detected in the mid-IR ($\simali$5--40$\mum$) spectra of
Galactic high-latitude clouds, normal star-forming galaxies, and
low-luminosity active galaxies. The presence of this hot dust
component cannot be explained by equilibrium heating. We propose that
the nanometer-size dust grains are transiently heated via
single-photon absorption. The most likely carrier are very small (size
\lax\ 10 \AA) carbonaceous (e.g., graphitic) grains. They account for $\simali$1.4\% of the total IR emission at $\simali$5--3000$\mum$, $\simali$1.5\% of the total carbon dust, and $\simali$0.4\% of the total dust mass.

\acknowledgments
We thank the anonymous referee for helpful suggestions which considerably 
improved the presentation of this paper.  
This work was supported by the National Key R\&D Program of China
(2016YFA0400702) and the National Science Foundation of China (11473002,
11721303).  Y.X. is supported by China Postdoctoral Science Foundation Grant 2016 M591007.
The Cornell Atlas of \spitzerirs Sources (CASSIS) is a product of the Infrared
Science Center at Cornell University, supported by NASA and JPL.

\clearpage
\begin{deluxetable}{lrrcclrl}
\tabletypesize\tiny
\tablecolumns{8}
\tablecaption{\label{tab:basic_T}
              Basic Parameters for the Sample of 
              Galaxies and Galactic 
              High-Latitude Clouds
              }
\tablewidth{0pt}
\tablehead{
\colhead{Source} & \colhead{R.A. (J2000)} & \colhead{Decl. (J2000)} & 
\colhead{Type} & \colhead{Distance}  & 
\colhead{$T^{h} $} & 
\colhead{$T^{w}$}  & 
\colhead{$T^{c}$}   \\ 
\colhead{}  & \colhead{(hh mm ss.s)} & \colhead{(dd mm ss.s)} & 
\colhead{}  & \colhead{(Mpc)} &  
\colhead{(\K)}  & 
\colhead{(\K)}  & 
\colhead{(\K)}  \\
\colhead{(1)} & \colhead{(2)} & \colhead{(3)} & \colhead{(4)} & \colhead{(5)}  & 
\colhead{(6)} & \colhead{(7)} & \colhead{(8)} 
}
\startdata
Mrk~~33 ................. & 10 32 31.82 & $+$54 24 02.5  &  H~II &  22.9     &   348  $\pm$      37.2  &        91  $\pm$       5.2  &        41   $\pm$      3.2 \\  
NGC~24 .................  & 00 09 56.37 & $-$24 57 51.2 &  H~II &   7.3     &   442  $\pm$      33.8  &        67  $\pm$       3.1  &        31   $\pm$      4.6 \\  
NGC~337 ...............   & 00 59 50.20 & $-$07 34 45.8 &  H~II &  22.4     &   427  $\pm$       7.2  &        69  $\pm$       0.1  &        31   $\pm$      0.2 \\  
NGC~628 ...............   & 01 36 41.60 & $+$15 47 00.0  &  H~II &   7.3     &   316  $\pm$       4.1  &        65  $\pm$       2.0  &        61   $\pm$     17.6 \\  
NGC~855 ...............   & 02 14 03.70 & $+$27 52 38.4  &  H~II &   9.6     &   687  $\pm$     255.3  &        67  $\pm$       1.0  &        32   $\pm$      0.9 \\  
NGC~925 ...............   & 02 27 17.25 & $+$33 34 41.6  &  H~II &   9.1     &   434  $\pm$       8.6  &        65  $\pm$       0.4  &        27   $\pm$      0.6 \\  
NGC~1482 .............   & 03 54 38.88 & $-$20 30 07.1  &  H~II &  23.2     &   455  $\pm$       1.7  &        70  $\pm$       0.2  &        37   $\pm$      0.6 \\  
NGC~2403 .............   & 07 36 49.95 & $+$65 36 03.5   &  H~II &   3.2     &   439  $\pm$       6.3  &        60  $\pm$       0.2  &        25   $\pm$      0.3 \\  
NGC~2798 .............   & 09 17 22.80 & $+$41 59 59.4   &  H~II &  26.2     &   325  $\pm$       2.4  &        87  $\pm$       0.1  &        46   $\pm$      0.1 \\  
NGC~2915 .............   & 09 26 10.03 & $-$76 37 32.2  &  H~II &   3.8     &   371  $\pm$       7.8  &        62  $\pm$       5.8  &        23   $\pm$      5.5 \\  
NGC~2976 .............   & 09 47 15.22 & $+$67 55 00.3 &    H~II &   3.6     &   313  $\pm$       2.8  &        67  $\pm$       0.3  &        27   $\pm$      0.4 \\  
NGC~3049 .............   & 09 54 49.59 & $+$09 16 18.1 &    H~II &  23.9     &   381  $\pm$      14.7  &        85  $\pm$       0.2  &        39   $\pm$      0.2 \\  
NGC~3184 .............   & 10 18 16.90 & $+$41 25 24.7 &    H~II &  11.1     &   354  $\pm$       7.7  &        74  $\pm$       0.3  &        32   $\pm$      0.5 \\  
NGC~3265 .............   & 10 31 06.80 & $+$28 47 45.6 &    H~II &  23.2     &   359  $\pm$      12.5  &        77  $\pm$       0.3  &        38   $\pm$      0.3 \\  
NGC~3351 .............   & 10 43 57.72 & $+$11 42 13.5 &    H~II &   9.3     &   349  $\pm$       2.3  &        82  $\pm$       0.1  &        39   $\pm$      0.1 \\  
NGC~3773 .............   & 11 38 12.98 & $+$12 06 45.8 &    H~II &  11.9     &   366  $\pm$      14.2  &        75  $\pm$       0.5  &        34   $\pm$      0.6 \\  
NGC~4254 .............   & 12 18 49.57 & $+$14 24 57.5 &    H~II &  16.6     &   375  $\pm$       2.7  &        72  $\pm$       0.2  &        35   $\pm$      0.2 \\  
NGC~4536 .............   & 12 34 27.03 & $+$02 11 16.5 &    H~II &  14.4     &   500  $\pm$       3.2  &        69  $\pm$       0.1  &        32   $\pm$      0.1 \\  
NGC~4559 .............   & 12 35 57.58 & $+$27 57 34.2 &    H~II &  10.3     &   386  $\pm$       9.5  &        63  $\pm$       1.8  &        27   $\pm$      1.5 \\  
NGC~4625 .............   & 12 41 52.68 & $+$41 16 26.9 &    H~II &   9.2     &   397  $\pm$      12.0  &        64  $\pm$       0.7  &        26   $\pm$      3.8 \\  
NGC~4631 .............   & 12 42 07.80 & $+$32 32 34.6 &    H~II &   8.1     &   473  $\pm$       2.1  &        66  $\pm$       0.3  &        34   $\pm$      0.6 \\  
NGC~5713 .............   & 14 40 11.38 & $-$00 17 24.2  &   H~II &  29.4     &   435  $\pm$       3.9  &        79  $\pm$       0.1  &        40   $\pm$      0.1 \\  
NGC~6946 .............   & 20 34 52.23 & $+$60 09 14.4  &   H~II &   6.8     &   394  $\pm$       1.5  &        74  $\pm$       0.0  &        39   $\pm$      0.1 \\  
NGC~7793 .............   & 23 57 49.84 & $-$32 35 27.1 &   H~II &   3.8     &   378  $\pm$       7.4  &        68  $\pm$       0.5  &        31   $\pm$      0.6 \\  
NGC~1097 .............   & 02 46 18.86 & $-$30 16 27.2 &   LINER &  17.1    &   360  $\pm$       0.5  &        81  $\pm$       0.0  &        41   $\pm$      0.0 \\  
NGC~1266 .............   & 03 16 00.71 & $-$02 25 36.9 &   LINER &  30.0    &   213  $\pm$       3.8  &        73  $\pm$       1.0  &        49   $\pm$      0.3 \\  
NGC~1512 .............   & 04 03 54.17 & $-$43 20 54.4 &   LINER &  11.8    &   407  $\pm$       5.9  &        76  $\pm$       0.6  &        37   $\pm$      0.5 \\  
NGC~1566 .............   & 04 20 00.33 & $-$54 56 16.6 &  Seyfert &  20.3   &   337  $\pm$       1.9  &       114  $\pm$       0.3  &        46   $\pm$      0.1 \\  
NGC~3198 .............   & 10 19 54.84 & $+$45 32 58.7 &   LINER &  13.7     &   235  $\pm$       0.9  &       107  $\pm$       5.9  &        65   $\pm$      0.1 \\  
NGC~3521 .............   & 11 05 48.58 & $-$00 02 07.3 &   LINER &  10.1    &   391  $\pm$       3.4  &       250  $\pm$       0.0  &        57   $\pm$      0.1 \\  
NGC~3621 .............   & 11 18 16.51 & $-$32 48 49.3 &   LINER &   6.6    &   426  $\pm$       4.1  &        65  $\pm$       0.2  &        29   $\pm$      0.4 \\  
NGC~3627 .............   & 11 20 15.04 & $+$12 59 29.0 &  Seyfert &   9.4    &   395  $\pm$       2.3  &        99  $\pm$       0.4  &        45   $\pm$      0.1 \\  
NGC~3938 .............   & 11 52 49.32 & $+$44 07 13.6 &   LINER &  13.3     &   314  $\pm$       1.8  &       250  $\pm$       0.0  &        63   $\pm$      0.2 \\  
NGC~4321 .............   & 12 22 54.87 & $+$15 49 19.2 &   LINER &  14.3     &   359  $\pm$       3.7  &        74  $\pm$       0.1  &        37   $\pm$      0.1 \\  
NGC~4450 .............   & 12 28 29.71 & $+$17 05 08.7 &   LINER &  16.6     &   500  $\pm$      11.7  &       128  $\pm$       2.3  &        40   $\pm$      1.0 \\  
NGC~4569 .............   & 12 36 49.76 & $+$13 09 45.5 & Seyfert &  16.6     &   356  $\pm$       2.4  &       101  $\pm$       0.2  &        41   $\pm$      0.1 \\  
NGC~4579 .............   & 12 37 43.53 & $+$11 49 03.8 & Seyfert &  16.6     &   361  $\pm$       3.5  &       124  $\pm$       1.0  &        43   $\pm$      0.3 \\  
NGC~4736 .............   & 12 50 53.15 & $+$41 07 14.4 &   LINER &   5.0     &  1500  $\pm$       0.0  &       214  $\pm$       0.2  &        48   $\pm$      0.0 \\  
NGC~4826 .............   & 12 56 43.59 & $+$21 40 58.0 & Seyfert &   5.0     &   435  $\pm$       1.2  &        76  $\pm$       0.1  &        38   $\pm$      0.1 \\  
NGC~5033 .............   & 13 13 27.32 & $+$36 35 35.2 & Seyfert &  14.8     &   390  $\pm$       2.2  &        88  $\pm$       0.7  &        45   $\pm$      0.2 \\  
NGC~5055 .............   & 13 15 49.35 & $+$42 01 45.7 &   LINER &   7.8     &   380  $\pm$       1.6  &       100  $\pm$       0.5  &        45   $\pm$      0.1 \\  
NGC~5194 .............   & 13 29 52.80 & $+$47 11 43.5 & Seyfert &   7.8     &   335  $\pm$       1.0  &        84  $\pm$       0.2  &        41   $\pm$      0.1 \\  
NGC~5195 .............   & 13 29 59.50 & $+$47 15 56.7 & Seyfert &   8.0     &   422  $\pm$       2.1  &       117  $\pm$       0.4  &        48   $\pm$      0.1 \\  
NGC~5866 .............   & 15 06 29.48 & $+$55 45 45.0 &   LINER &  15.1     &   864  $\pm$      47.4  &       193  $\pm$       3.1  &        44   $\pm$      0.2 \\  
NGC~7331 .............   & 22 37 04.15 & $+$34 24 55.3 &   LINER &  14.5     &   467  $\pm$       4.7  &       131  $\pm$       1.5  &        48   $\pm$      0.2 \\  
NGC~7552 .............   & 23 16 10.83 & $-$42 35 05.5 &   LINER &  21.0    &   339  $\pm$       1.1  &        91  $\pm$       0.1  &        48   $\pm$      0.1 \\  
DCld~300.2-16.9 (A) & 11 54 23.10 & $-$79 31 42.0 & HLC  & 70$\pm$15E-6        &   500  $\pm$       6.9  &       102  $\pm$       0.5  &        45   $\pm$      0.4 \\  
DCld~300.2-16.9 (B) & 11 52 08.30 & $-$79 09 33.0 & HLC  & 70$\pm$15E-6        &   460  $\pm$       3.6  &        84  $\pm$       0.6  &        45   $\pm$      0.6 \\  
DCld~300.2-16.9 (C) & 11 48 24.40 & $-$79 18 00.0 & HLC  & 70$\pm$15E-6        &   371  $\pm$       2.4  &        95  $\pm$       0.5  &        45   $\pm$      0.5 \\  
DCld~300.2-16.9 (D) & 11 55 33.80 & $-$79 20 54.0 & HLC  & 70$\pm$15E-6        &   394  $\pm$       4.1  &       106  $\pm$       0.6  &        48   $\pm$      0.4 \\  
\enddata
\end{deluxetable}

\begin{deluxetable}{lccrrrrcc}
\tabletypesize\tiny
\tablecolumns{9}
\tablecaption{{\it Herschel}\  Photometry of GOALS Galaxies Used in Our Analysis\label{tab:goals}} 
\tablewidth{0pt}
\tablehead{
\colhead{Source}                       &  \colhead{Redshift}                        & \colhead{log$ (\frac {L_{\mathrm {IR}}}{L_{\odot}}$)} & 
\colhead{$F_{\mathrm{70}}$}   &  \colhead{ $F_{\mathrm{100}}$}  & \colhead{$F_{\mathrm{160}}$} &
\colhead{$F_{\mathrm{250}}$} &  \colhead{$F_{\mathrm{350}}$}  & \colhead{$F_{\mathrm {500}}$} \\
\colhead{} & \colhead{} & \colhead{($L_{\odot}$)} & \colhead{(Jy) } & \colhead{(Jy)} & \colhead{(Jy)} & \colhead{(Jy)} & \colhead{(Jy)} & \colhead{(Jy)}  \\
\colhead{(1) } & \colhead{(2)} & \colhead{(3)} & \colhead{(4)} & \colhead{(5)} & \colhead{(6)} & \colhead{(7)} & \colhead{(8)} & \colhead{(9)} 
}
\startdata
ESO~286-G035         &     0.01736 &   11.20 &    9.50$\pm$0.44     &   12.80$\pm$0.58     &   10.40$\pm$0.47     &    3.99$\pm$0.24     &    1.47$\pm$0.09     &    0.40$\pm$0.03    \\
ESO~319-G022         &     0.01635 &   11.12 &    9.09$\pm$0.41     &   10.50$\pm$0.48     &    7.79$\pm$0.34     &    3.09$\pm$0.19     &    1.14$\pm$0.07     &    0.32$\pm$0.02    \\
ESO~353-G020         &     0.01600 &   11.06 &   10.60$\pm$0.49     &   16.40$\pm$0.75     &   15.30$\pm$0.68     &    6.57$\pm$0.41     &    2.62$\pm$0.16     &    0.88$\pm$0.06    \\
ESO~557-G002         &     0.02130 &   11.25 &    8.70$\pm$0.34     &   10.30$\pm$0.38     &    7.69$\pm$0.28     &    3.01$\pm$0.14     &    1.16$\pm$0.06     &    0.35$\pm$0.02    \\
IC~0860              &     0.01116 &   11.14 &   19.70$\pm$0.90     &   18.10$\pm$0.82     &   10.70$\pm$0.47     &    3.78$\pm$0.23     &    1.52$\pm$0.09     &    0.50$\pm$0.03    \\
IC~2810              &     0.03400 &   11.64 &    8.46$\pm$0.28     &   11.40$\pm$0.38     &    9.35$\pm$0.30     &    3.82$\pm$0.25     &    1.61$\pm$0.11     &    0.54$\pm$0.04    \\
IC~4734              &     0.01561 &   11.35 &   19.10$\pm$0.88     &   25.40$\pm$1.16     &   20.50$\pm$0.92     &    8.45$\pm$0.53     &    3.26$\pm$0.20     &    1.05$\pm$0.07    \\
IRAS~F01364-1042     &     0.04825 &   11.85 &    7.40$\pm$0.33     &    6.85$\pm$0.31     &    3.99$\pm$0.17     &    1.34$\pm$0.08     &    0.52$\pm$0.03     &    0.16$\pm$0.01    \\
IRAS~F03514+1546     &     0.02222 &   11.20 &    6.67$\pm$0.33     &    8.85$\pm$0.44     &    6.95$\pm$0.35     &    3.07$\pm$0.19     &    1.13$\pm$0.07     &    0.36$\pm$0.03    \\
IRAS~F06076-2139     &     0.03745 &   11.65 &    7.55$\pm$0.34     &    8.41$\pm$0.37     &    5.66$\pm$0.24     &    2.16$\pm$0.13     &    0.86$\pm$0.05     &    0.27$\pm$0.02    \\
IRAS~F06592-6313     &     0.02296 &   11.24 &    6.90$\pm$0.31     &    7.53$\pm$0.34     &    5.03$\pm$0.22     &    1.84$\pm$0.11     &    0.76$\pm$0.05     &    0.22$\pm$0.02    \\
IRAS~F10565+2448     &     0.04310 &   12.08 &   14.30$\pm$0.64     &   15.80$\pm$0.71     &   10.50$\pm$0.46     &    3.64$\pm$0.22     &    1.34$\pm$0.08     &    0.38$\pm$0.02    \\
IRAS~F14179+4927     &     0.02574 &   11.39 &    6.51$\pm$0.33     &    7.45$\pm$0.37     &    4.90$\pm$0.25     &    1.79$\pm$0.12     &    0.69$\pm$0.05     &    0.18$\pm$0.02    \\
IRAS~F16284+0411     &     0.02449 &   11.45 &    9.15$\pm$0.41     &   12.50$\pm$0.55     &   10.00$\pm$0.44     &    3.62$\pm$0.24     &    1.42$\pm$0.10     &    0.49$\pm$0.04    \\
IRAS~F17132+5313     &     0.05094 &   11.96 &    6.21$\pm$0.31     &    6.98$\pm$0.35     &    5.47$\pm$0.24     &    2.01$\pm$0.12     &    0.78$\pm$0.05     &    0.23$\pm$0.02    \\
IRAS~F17207-0014     &     0.04281 &   12.46 &   38.10$\pm$1.74     &   37.90$\pm$1.72     &   23.10$\pm$1.02     &    7.96$\pm$0.49     &    2.91$\pm$0.18     &    0.89$\pm$0.05    \\
IRAS~F18293-3413     &     0.01818 &   11.88 &   45.70$\pm$2.11     &   59.10$\pm$2.71     &   45.80$\pm$2.07     &   17.20$\pm$1.07     &    6.45$\pm$0.40     &    1.99$\pm$0.12    \\
IRAS~F22491-1808     &     0.07776 &   12.20 &    5.64$\pm$0.25     &    4.98$\pm$0.22     &    2.72$\pm$0.12     &    0.84$\pm$0.05     &    0.31$\pm$0.02     &    0.08$\pm$0.01    \\
MCG~02-04-025        &     0.03123 &   11.69 &   11.70$\pm$0.53     &   11.30$\pm$0.51     &    6.91$\pm$0.31     &    2.13$\pm$0.13     &    0.77$\pm$0.05     &    0.23$\pm$0.02    \\
MCG~02-20-003        &     0.01625 &   11.13 &   10.60$\pm$0.48     &   13.10$\pm$0.59     &    9.97$\pm$0.44     &    4.07$\pm$0.25     &    1.56$\pm$0.10     &    0.50$\pm$0.03    \\
MCG~03-04-014        &     0.03349 &   11.65 &    9.00$\pm$0.41     &   11.40$\pm$0.51     &    8.76$\pm$0.39     &    3.08$\pm$0.19     &    1.18$\pm$0.07     &    0.35$\pm$0.02    \\
MCG~05-12-006        &     0.01875 &   11.17 &    9.17$\pm$0.41     &   10.40$\pm$0.46     &    6.95$\pm$0.30     &    2.45$\pm$0.15     &    0.91$\pm$0.06     &    0.27$\pm$0.02    \\
UGC~01385            &     0.01875 &   11.05 &    6.77$\pm$0.30     &    8.27$\pm$0.36     &    6.29$\pm$0.25     &    2.54$\pm$0.13     &    1.03$\pm$0.06     &    0.28$\pm$0.03    \\
UGC~01845            &     0.01561 &   11.12 &   13.10$\pm$0.59     &   17.50$\pm$0.78     &   13.80$\pm$0.60     &    5.04$\pm$0.30     &    1.97$\pm$0.12     &    0.56$\pm$0.03    \\
UGC~12150            &     0.02139 &   11.35 &   10.70$\pm$0.48     &   14.70$\pm$0.66     &   12.30$\pm$0.54     &    5.09$\pm$0.31     &    2.03$\pm$0.12     &    0.61$\pm$0.04    \\
VII~Zw~031           &     0.05367 &   11.99 &    7.49$\pm$0.34     &   10.10$\pm$0.45     &    8.06$\pm$0.35     &    3.13$\pm$0.19     &    1.17$\pm$0.07     &    0.35$\pm$0.02    \\
VV~250a              &     0.03107 &   11.81 &   11.70$\pm$0.59     &   11.20$\pm$0.56     &    6.73$\pm$0.34     &    2.43$\pm$0.16     &    0.88$\pm$0.06     &    0.27$\pm$0.02    \\
UGC~03608            &     0.02135 &   11.33 &    8.88$\pm$0.44     &   10.60$\pm$0.53     &    8.32$\pm$0.42     &    3.59$\pm$0.22     &    1.36$\pm$0.08     &    0.43$\pm$0.03    \\
\enddata
\end{deluxetable}


%

\end{document}